\documentclass{article}
\usepackage{spconf,amsmath,graphicx}
\usepackage{booktabs}

\title{Enhancing Twitter Bot Detection via Multimodal Invariant Representations}
\name{Jibing Gong, Jiquan Peng, Jin Qu, ShuYing Du, Kaiyu Wang\thanks{Thanks to CCF-Zhipu AI Large Model Fund (CCF-Zhipu202307), CIPSC-SMP-Zhipu.AI Large Model Cross-Disciplinary Fund and Innovation Capability Improvement Plan Project of Hebei Province (22567626H).}}
\address{School of Information Science and Engineering, Yanshan University\\
The Key Laboratory for Computer Virtual Technology and System Integration of Hebei Province\\
}
\begin{document}
%\ninept
%
\maketitle
\begin{abstract}
Detecting Twitter Bots is crucial for maintaining the integrity of online discourse, safeguarding democratic processes, and preventing the spread of malicious propaganda. However, advanced Twitter Bots today often employ sophisticated feature manipulation and account farming techniques to blend seamlessly with genuine user interactions, posing significant challenges to existing detection models. In response to these challenges, this paper proposes a novel Twitter Bot Detection framework called BotSAI. This framework enhances the consistency of multimodal user features, accurately characterizing various modalities to distinguish between real users and bots. Specifically, the architecture integrates information from users, textual content, and heterogeneous network topologies, leveraging customized encoders to obtain comprehensive user feature representations. The heterogeneous network encoder efficiently aggregates information from neighboring nodes through oversampling techniques and local relationship transformers. Subsequently, a multi-channel representation mechanism maps user representations into invariant and specific subspaces, enhancing the feature vectors. Finally, a self-attention mechanism is introduced to integrate and refine the enhanced user representations, enabling efficient information interaction. Extensive experiments demonstrate that BotSAI outperforms existing state-of-the-art methods on two major Twitter Bot Detection benchmarks, exhibiting superior performance. Additionally, systematic experiments reveal the impact of different social relationships on detection accuracy, providing novel insights for the identification of social bots.
\end{abstract}
\section{Introduction}
\label{sec:intro}

As a globally popular online social media platform, Twitter has become a multifaceted site where people share life events and engage in discussions on trending topics. However, it also harbors a substantial number of automated Twitter Bot accounts. These bots pose a threat and challenge to the structure of online communities by disseminating misinformation~\cite{cresci2023demystifying}, interfering in national elections~\cite{yang2021covid} , engaging in privacy attacks~\cite{varol2017online} , and promoting extremist ideologies~\cite{he2024dynamicity} . Therefore, effective Twitter Bot Detection methods are urgently needed to mitigate their negative impact.

An arms race has begun between the creation and detection of Twitter Bots. Early simple Twitter Bots, which were created using random information, influenced public opinion by posting repetitive and malicious content. Twitter Bot Detection models could easily identify these bots by feeding user attributes extracted from metadata and user activity timelines into traditional classifiers~\cite{yang2020scalable,mazza2019rtbust}. In response, Twitter Bots began deliberately altering user metadata to evade feature-based detection methods. Researchers countered this by employing Natural Language Processing (NLP) techniques, such as Recurrent Neural Networks (RNNs) and pre-trained models~\cite{d2015real,wei2019twitter,diaz2020integrated,heidari2020using}, to encode user tweets and detect malicious content based on semantic recognition. However, Twitter Bots then started obfuscating false tweets with tweets stolen from real users. Researchers tackled this by leveraging the graphical structure of the Twittersphere, which is composed of social relationships among Twitter users. They used Graph Neural Networks (GNNs) like Graph Convolutional Networks (GCNs)~\cite{ali2019detect}, Relational Graph Convolutional Networks (RGCNs)~\cite{feng2021botrgcn}, and Relational Graph Transformers (RGTs)~\cite{feng2022heterogeneity} for graph node classification to detect bots. Graph-based methods outperform text-based methods in detection performance and exhibit better generalization capabilities~\cite{feng2022twibot}. Today, advanced bots create accounts using real users' descriptive information and employ account farming techniques for camouflage, such as following and commenting on real users and continuously engaging in real user interactions to hide themselves and expand their influence. This represents a more sophisticated level of deception compared to earlier methods. 
\begin{figure}[htbp]
  \centering
  \includegraphics[width=\linewidth]{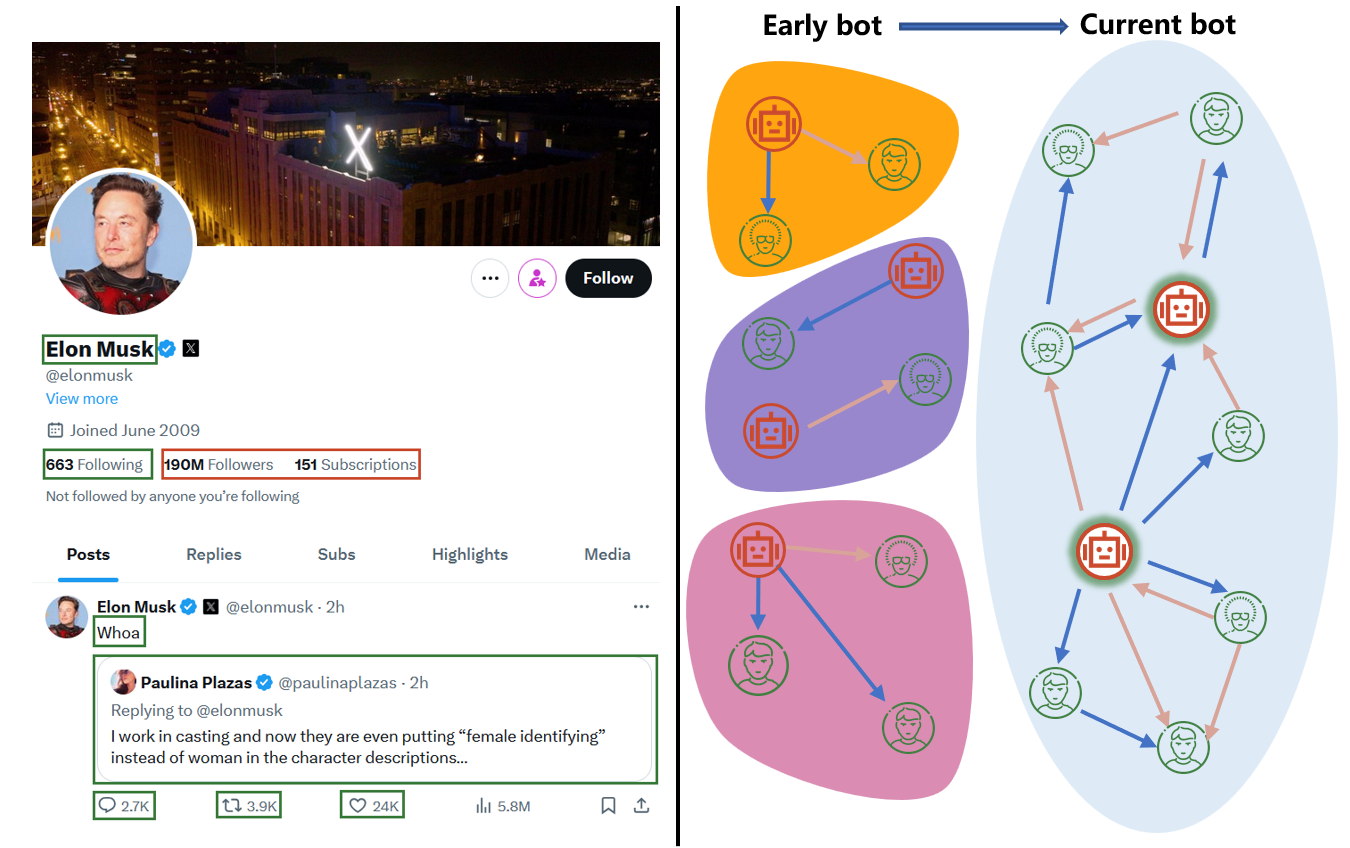}
  \caption{
  Twitter bots can manipulate messages with deep social network camouflage.}
  \label{fig:intro}
\end{figure}
As illustrated in Figure~\ref{fig:intro}, the left half shows the real user information that Twitter Bots can use for disguise, with green boxes indicating easily replicable active behaviors and red boxes indicating harder-to-replicate passive behaviors. The right half of the figure shows the simple social network of early Twitter Bots, characterized by fewer user nodes and simpler relationships. In contrast, today's bots, through account farming, expand their social networks to include more user nodes and complex relationships. Despite significant progress in Twitter Bot Detection research, current methods still struggle against the sophisticated disguises of modern bots. Existing detection approaches face three key challenges:
1) Key datasets like MGTAB~\cite{shi2023mgtab} and TwiBot-20~\cite{feng2022twibot} exhibit a significant imbalance between bots and real users, leading to biased predictions favoring the majority class. Previous methods have failed to mitigate this issue effectively.
2) Twitter users exhibit diverse social relationships. Most graph-based methods only consider follower and friend relationships, overlooking the adaptive capture of useful information from various relationship types. The impact of using multiple relationships on detection accuracy remains underexplored.
3) Utilizing multimodal information is essential for modeling user characteristics. However, the inherent heterogeneity between modalities can negatively impact detection performance. Developing reliable methods to leverage multimodal information is crucial for improving detection outcomes.

To gain an advantage in the arms race of bot detection, we propose a novel Twitter Bot Detection framework called BotSAI. This framework employs a hybrid architecture with invariant and specific subspace awareness. Specifically, it extracts features from Twitter users' metadata, textual content, and heterogeneous network topology using redesigned modality-specific encoders. A local relationship graph transformer based on graph information synthesizes minority class samples using oversampling strategies and leverages multiple social relationships for feature extraction. Subsequently, multi-channel representers map each modality's feature vectors to invariant and specific representations, learning commonalities within modalities and highlighting inter-modal characteristics. Multi-head self-attention is then used for feature interaction. The resulting feature vectors, rich in effective information, address the social bot detection task. In summary, the main contributions of this paper are:
\begin{itemize}
\item By utilizing Twitter users' metadata, textual content, and heterogeneous network topology, and ensuring data balance in the Twittersphere while selecting efficient social relationships, our bot detection model can robustly identify subtle differences between real users and bots.

\item We introduce a new Twitter Bot Detection framework that models comprehensive user representations, enhances integration by mapping multimodal information to invariant and specific subspaces, and effectively captures differences between real users and bots. This is an end-to-end detection method.

\item We conducted extensive experiments to evaluate the model's performance and compared it with state-of-the-art methods. The results show that BotSAI performs excellently across all benchmarks. Systematic experiments also explored the impact of different social relationships on detection accuracy.
\end{itemize}

\section{Approach}
\label{sec:Approach}
\subsection{Task Setup}
The aim of this paper is to detect Twitter bots through the utilization of multi-modal signals. The detection task is based on three distinct feature sequences: graphical modal features $G$, textual modal features $T$, and metadata features $M$.
The heterogeneous graph composed of social relationships of Twitter users is defined as $G=G\left(U, E, \varphi, R^{e}\right)$, where $U=\left\{u_{1}, u_{2}, \ldots, u_{n}\right\}$ is the set of users, $E$ is the edge set, $\varphi:E\to R^{e}$ represents the relational mapping function, and $R^{e}$ is the set of relational types.
Twitter user $u_{v} $'s neighbors can be exported from $G$ as $N_{v} =\{u_{v,w} \}_{w=1}^{W_{v} } $, where $W_{v}$ is the number of neighbors.
The text modality contains the a description information of user $u_{v}$ denoted as $B_{v} =\{t_{v}^{a}\}_{a=1}^{Q_{a}} $ and the b tweet information denoted as $T_{v} =\{t_{v}^{b}\}_{b=1}^{Q_{b}} $, where $Q_{a}$ and $Q_{a}$ denote the total amount. The metadata modal input is denoted as $m_{v}$. The goal is to find a detection function $f:f(u_{v} )\to \hat{y} \in \{0,1\}$ such that $\hat{y}$ approximates the true label value $y$ such that the prediction accuracy is maximized. The BotSAI framework process can be divided into two main stages: multimodal information representation learning and multimodal information interaction stage.

\subsection{Overview}
\begin{figure*}[htbp]  
\centering  
\includegraphics[width=\linewidth]{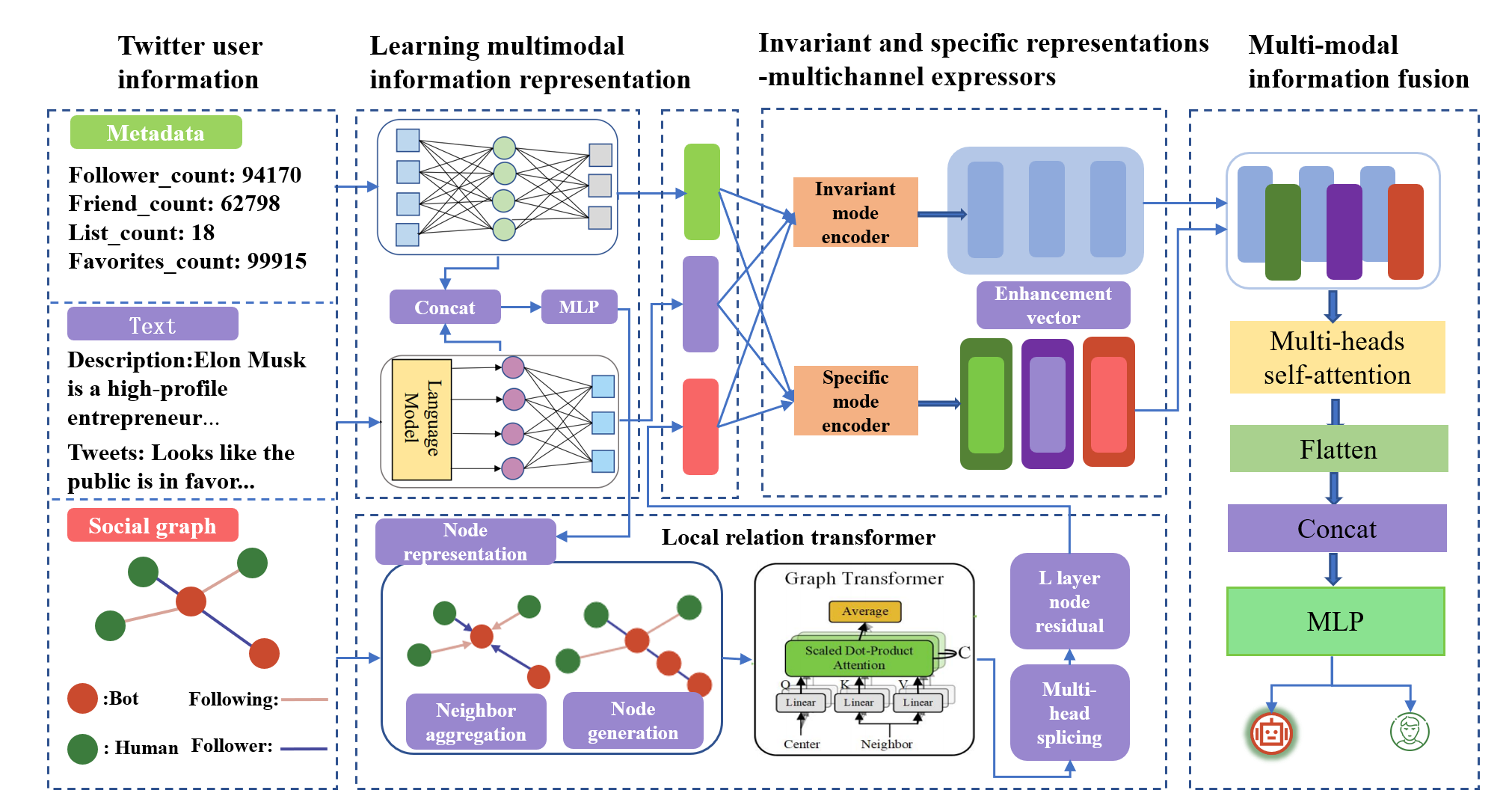}  
\caption{Overview of the Twitter bot detection framework BotSAI}
\label{fig:overview}  
\end{figure*}
Figure~\ref{fig:overview} illustrates the Twitter social bot detection framework, BotSAI, which integrates multi-modal user information encoding with the capability for modal invariant mapping across distinct subspaces. The framework begins with the extraction of features from user metadata, text content, and heterogeneous network topology. This is achieved through dedicated encoders: a redesigned metadata encoder, text encoder, and graph encoder, each designed to capture a comprehensive representation of Twitter users from various perspectives. Following feature extraction, each feature is mapped to both modal invariant and modal-specific subspaces. A multi-head self-attention mechanism is then applied to the enhanced eigenvector representations, enabling each vector to assimilate information from cross-modal and cross-subspace representations. This fusion approach allows for the induction of latent information from other representations, thereby enhancing the overall effectiveness of the Twitter bot detection task. Ultimately, the output fusion feature vectors are concatenated to form joint vectors, which are employed in detection tasks and model parameter learning.

\subsection{Multimodal Representation Learning}
Twitter bot operators often create obfuscated accounts by using web crawlers to harvest real users' personal information and tweets, making it challenging to identify bots based solely on text information. Incorporating the social graph of Twitter users significantly enhances the distinction between bots and genuine users, thereby improving detection accuracy. However, relying solely on social graph information can overly smooth the crucial textual content of Twitter accounts. Since text is the primary medium through which bot accounts disseminate false information, its detection efficacy is suboptimal when evaluated experimentally. To address these challenges, this paper proposes a comprehensive user representation model that integrates metadata, text content, and heterogeneous network topology information of Twitter users.
\subsubsection{Metadata encoder}
This study examines metadata from user profiles, focusing on five numerical attributes (followers count, following count, status count, active days, and screen name length) and three categorical attributes (account protection status, verification status, and default profile image status) based on their availability and simplicity. These feature embeddings are denoted as \( \mathbf{m}_v \). Following Z-Score normalization, two layers of a multilayer perceptron (MLP) are applied to learn the feature representation of user \( u_v \)'s metadata, resulting in \( \mathbf{x}_{v}^{m} \).
\begin{equation}
  x_{v}^{M}=MLP(m_{v}) 
\end{equation}
\subsubsection{Text encoder}
In this study, a pre-trained language model was employed to encode the tweets and descriptive information of Twitter users within the dataset. For consistency and comparability, the same RoBERTa model [28] utilized by other baseline models was used as the encoder for the text module. The procedure is outlined as follows:
\begin{equation}
  \mathbf{t}_{v}=\frac{1}{n} \sum_{a=1}^{n} \frac{1}{l_{a}} \sum_{b=1}^{l_{n}} \mathrm{LM}\left(\mathbf{t}_{v}\right)^{a, b}
\end{equation}
In this context, let \( LM \) denote the pre-trained language model encoder, \( t_v \) represent the textual information of user \( u_v \), \( a \) signify a specific tweet within the tweet data, \( b \) indicate an attribute within the descriptive information, \( n \) represent the total number of text contents, and \( l_a \) denote the length of the input text sequence in Article \( a \). The features extracted from user \( u_v \)'s descriptive information are denoted as \( t_{v}^{d} \), while the features extracted from user \( u_v \)'s tweet information are denoted as \( t_{v}^{a} \). These features are subsequently input into a two-layer Multi-Layer Perceptron (MLP) to obtain the low-dimensional text representation \( x_{v}^{T} \) for user \( u_v \).

\subsubsection{Graph encoder}
To account for the diverse relationships present in social networks, this study constructs a heterogeneous information network to capture the intricate social interactions among Twitter users. The initial characteristics of Twitter users, representing the node information in the social relationship network graph, are derived from user metadata, personal descriptions, and tweet data. This comprehensive approach facilitates the creation of a robust user node embedding. The original characteristics of the nodes are expressed as follows:
\begin{equation}
  h_v=W_D\cdot (x_v^M\parallel x_v^T)+b_D
\end{equation}
Due to the imbalance between the number of bot accounts and real user accounts in the dataset, the domain node features of the user nodes in the social graph are mapped to the feature space using a graph neural network (GNN). To address the dataset imbalance, oversampling is employed to generate node samples for underrepresented classes in the feature space. Initially, a 2-hop neighborhood aggregation is utilized. The neighborhood aggregation information is expressed as follows. Given that bot accounts may conceal themselves through interactions with multiple real users within the Twitter social relationship network, to prevent excessive smoothing and concealment of bot accounts through 2-hop neighborhood aggregation, both the and the connection are used as the embedded representation of user nodes.
\begin{equation}
  x_v=[q_v\parallel h_v]
\end{equation}
This method of integrating the original node information with the aggregated neighbor information preserves the inherent characteristics of the nodes while making the embedding of bot nodes similar to that of human nodes. Let \( x_v \) denote the embedded representation of user node \( u_v \). Minority class nodes are generated using the synthetic node generator, with the number of generated nodes controlled by the hyperparameter \( \omega \) (oversampling scale). For a minority class node \( u_v \), let \( \Gamma (\cdot ) \) represent the set of \( k \) neighbors of \( u_v \) as measured by Euclidean distance in the feature space. A random node \( u_w \) from \( \Gamma (u_v ) \) with the same label as \( u_v \) is selected, and a random point on the line segment between \( u_v \) and \( u_w \) is chosen as \( u_k \).
\begin{equation}
  x_k=(1-\delta )\cdot x_v+\delta \cdot x_W
\end{equation}
Here, \( x_k \) represents the embedding of the generated virtual node in the feature space. The variable \( \delta \) is a random variable uniformly distributed between 0 and 1. The virtual node \( x_k \) shares the same label as \( x_v \) and \( x_w \). Consequently, this method produces a small number of uniformly distributed class samples.

The heterogeneity of influence, global relationships, and local relationships is prevalent in Twitter social networks. The semantic attention network of the RGT~\cite{feng2022heterogeneity} evaluates the importance of each social relationship from a global perspective, assigning equal weight to different neighbors of Twitter users within the same social relationship. This approach overlooks the varying influence of different neighbors under the same social relationship, failing to fully capture user individuality. Inspired by the success of Transformers~\cite{vaswani2017attention} in natural language processing, this paper proposes a Local Relational Graph Converter, which employs a GNN architecture incorporating Transformers to operate on heterogeneous information networks, thereby modeling influence heterogeneity and local relational heterogeneity. A multi-head attention mechanism is utilized to aggregate and calculate the influence between the central node and neighboring nodes in the heterogeneous social relationship graph. Initially, linear transformations are applied to the central node feature \( x_v^{l-1} \) and the neighbor node feature \( x_u^{l-1} \). The resulting query, key, and value of the \( l \)-layer GNN based on the \( i \)-th attention head of relation \( r \) are expressed as follows:
\begin{equation}
\begin{aligned}
    Q_{v}^{(l), r, i}=W_{Q}^{(l), r, i} x_{v}^{(l-1)}+b_{Q}^{(l), r, i} \\
    K_{u}^{(l), r, i}=W_{K}^{(l), r, i} x_{u}^{(l-1)}+b_{K}^{(l), r, i} \\
    V_{u}^{(l), r, i}=W_{V}^{(l), r, i} x_{u}^{(l-1)}+b_{V}^{(l), r, i}
\end{aligned}
\end{equation}
Here, \( W_{*}^{(l), r, i} \) and \( b_{*}^{(l), r, i}\) are learnable parameters. \( Q \), \( K \), and \( V \) serve as the query, key, and value of the Transformer, respectively. The attention score of the neighbor node to the central node is then calculated based on the obtained \( Q \), \( K \), and \( V \), as follows:
\begin{equation}
  e_{v u}^{(l), r, i}=\frac{Q_{v}^{(l), r, i} \cdot K_{u}^{(l), r, i}}{\sqrt{d_{k}}}
\end{equation}
Here, \( e_{v u}^{(l), r, i} \) represents the importance score of the central node \( v \) with respect to the neighbor \( u \) under the deterministic relation \( r \). The dimension of the key vector \( d_k \) is used to scale the results of the query and key dot product operation. Subsequently, the attention scores of all neighbor nodes are normalized using the softmax function to obtain the attention weights, as illustrated below:
\begin{equation}
  \alpha_{v u}^{(l), r, i}=\operatorname{softmax}\left(e_{v u}^{(l), r, i}\right)=\frac{\exp \left(e_{v u}^{(l), r, i}\right)}{\sum_{k \in \mathrm{N}_{r}(v)} \exp \left(e_{v k}^{(l), r, i}\right)}
\end{equation}
Here, \( \alpha_{v u}^{(l), r, i} \) denotes the attention weight of the central node \( v \) concerning the neighbor node \( u \) under a specific relation \( r \). This allows for the determination of the attention weight between the central node and its neighboring nodes, classified by the relation. The recombination of attention weights reflects the relative importance of each neighboring node to the central node, accounting for the heterogeneity in the local relationship between the central user node and the neighbor user node. The output of the \( i \)-th attention head of the central node \( v \) is obtained by performing a weighted summation of the neighbor node values using the normalized attention weights, as follows:
\begin{equation}
  z_v^{(l),i}=\sum_{r\in R }^{} \sum_{u\in N_r(v) }^{} \alpha_{v u}^{(l), r, i}V_{u}^{(l), r, i}
\end{equation}
Here, \( z_v^{(l),i} \) denotes the output of the \( i \)-th attention head at the \( l \)-th layer in the GNN for the central user node \( v \). The outputs from each attention head are concatenated and subsequently mapped back to the original dimension through a linear transformation.
\begin{equation}
  H_{v}^{(l)}=W^{(l), O}\left[z_{v}^{(l), 1}\left\|z_{v}^{(l), 2}\right\| \ldots \| z_{v}^{(l), h}\right]
\end{equation}
Here, \( H_{v}^{(l)} \) denotes the final representation of the central user node \( v \), \( W^{(l), O} \) represents the learnable output linear transformation matrix for layer \( l \), and \(\parallel \) indicates the column concatenation operation.

In the cross-layer representation of user nodes, residual connections within the graph neural network are incorporated into the hidden representations of user nodes to facilitate smooth representation learning. Specifically, this approach is expressed as follows:
\begin{equation}
  x_v^{(l)}=\beta H_v^{(l)}+(1-\beta)x_v^{(l-1)}
\end{equation}
Here, \( \beta \) is a learning parameter used to control the weighted ratio between the feature \( H_v^{(l)} \) after multi-head attention transformation and the original feature \( x_v^{(l-1)} \). Following the aggregation update, the final representation is extracted as \( x_v^{G} \).

\subsection{Multimodal Information Interaction}
Initially, invariant and specific representation enhancement is applied to the feature vectors derived from the metadata, text, and graphical modes. The first representation is modal-invariant and aims to reduce the discrepancy between modes while capturing the underlying common features. This involves mapping all multimodal information from Twitter users into a shared subspace with distribution alignment. Despite originating from different sources, these signals reflect the common intentions and goals of users, which are crucial for a comprehensive assessment of the authenticity of Twitter users.

The second representation is mode-specific, where BotSAI learns the unique features inherent to each mode. Each mode possesses its own distinct characteristics, such as user-specific style information. Although these specific details may not be directly usable for bot detection and might be considered noise, they are essential for the task. For instance, the expression style of the user's tweets and personal interests must be taken into account. Learning these mode-specific features complements the common underlying features captured in the invariant space, thereby providing a comprehensive multimodal representation of the user.

To learn these two subspaces, this study applies distributed similarity losses to the invariant subspaces and uses orthogonal losses for the specific subspaces. Additionally, reconstruction losses are used to ensure that the hidden representation captures the details specific to each mode. The fusion-enhanced user representation is then employed for Twitter bot detection.

\subsubsection{Multichannel Indicator}
To fully utilize the multimodal information of Twitter users, and inspired by the successful application of invariant subspace and specific subspace architectures in graph classification and sentiment classification, a multichannel representation is employed. This approach projects each user's vector representation from the metadata, text, and graph modal encoders into two distinct spaces. Initially, each modal feature vector is projected into a specific subspace to capture unique features pertinent to each mode. The encoding functions used for this process are as follows:
\begin{equation}
\begin{aligned}
h_{s}^{G}=E_{s}\left(x_{v}^{G} ; \theta_{s}^{G}\right) \\
h_{s}^{T}=E_{s}\left(x_{v}^{T} ; \theta_{s}^{T}\right) \\
h_{s}^{M}=E_{s}\left(x_{v}^{M} ; \theta_{s}^{M}\right)
\end{aligned}
\end{equation}
This method effectively separates the unique information contained in each mode, allowing the characteristics of each mode to be fully expressed in its distinct feature space. Subsequently, the eigenvectors of each mode are projected into the invariant subspace. The following encoding functions are used:
\begin{equation}
\begin{aligned}
h_{i}^{G}=E_{i}\left(x_{v}^{G} ; \theta_{i}\right) \\
h_{i}^{T}=E_{i}\left(x_{v}^{T} ; \theta_{i}\right) \\
h_{i}^{M}=E_{i}\left(x_{v}^{M} ; \theta_{i}\right)
\end{aligned}
\end{equation}
This constraint mitigates the structural differences between various modal data, facilitating the integration of multimodal information within a common feature space and establishing a consistent feature representation across different modes. In the formula, \( E \) denotes the feedforward neural layer, where \( \theta_{s}^{G} \), \( \theta_{s}^{T} \), and \( \theta_{s}^{M} \) are distinct parameters assigned to each mode in \( E_{i} \), and \( \theta_{i} \) is a shared parameter across all three modes. Enhancing features through multi-channel representators provides a comprehensive perspective essential for effective fusion.

\subsubsection{Multimodal Feature Fusion}
Once the feature representation of each mode is projected into specific and invariant subspaces, the internal correlation between each representation is captured through multi-head self-attention. The resulting six transformed modal vectors are then concatenated for the Twitter bot detection task. In this study, the matrix \( M=[h_G^i,h_T^i,h_M^i,h_G^s,h_T^s,h_M^s] \) is derived using the enhanced eigenvector representation of series projection into invariant and specific subspaces. The queries, keys, and values for the attention calculation are subsequently obtained through linear transformations:
\begin{equation}
\begin{aligned}
Q_i=W_i^QM+b_i^Q\\
K_i=W_i^KM+b_i^K\\
V_i=W_i^VM+b_i^V
\end{aligned}
\end{equation}
Where \( W_i^Q \), \( W_i^K \), and \( W_i^V \), and \( b_i^Q \), \( b_i^K \), and \( b_i^V \) are learnable parameters. Substituting the resulting query, key, and value into the attention module of the scaled dot-product function yields the output of the \( i \)th attention head as follows:
\begin{equation}
  head_i=Attention(Q_i,K_i,V_i)=softmax(\frac{Q_iK_{i}^{T} }{\sqrt{d_h} } )V_i
\end{equation}
Subsequently, the outputs of multiple self-attention heads are concatenated and linearly transformed to obtain the multi-head self-attention output:
\begin{equation}
  \bar{M}=MultiHead(M;\theta ^{att})=(head_1\oplus \cdots head_n )W_o 
\end{equation}
Let \( W_o \) denote a learnable parameter, while \( \oplus \) signifies the series and \( \theta ^{att}=\{W^q,W^k,W^v,W^o\} \). The column vectors within the concatenation matrix \( \bar{M} \) form a composite vector, \( h_{\text{out}} \). Subsequently, the task detection function is applied to \( h_{\text{out}} \) to derive the final label.
\begin{equation}
  \hat{y} =D(h_{out};\theta_{out} )
\end{equation}

\subsection{Optimization}
The optimization of the model's predictive framework is achieved through the minimization of specific loss functions.
\begin{equation}
  L=L_{task}+\alpha L_{sim}+\beta L_{diff}+\gamma L_{recon}
\end{equation}
Here, \(\alpha\), \(\beta\), and \(\gamma\) are hyperparameters that regulate the contribution of each component loss to the overall loss. These component losses are instrumental in ensuring the model achieves the desired subspace properties.

\subsubsection{Similarity Loss $L_{sim}$}
Similarity loss is employed to characterize the invariant representations of various modes. By minimizing similarity loss, one can reduce the discrepancy between shared representations across different modes. This approach, when applied to invariant subspaces, facilitates the alignment of cross-modal features. In this study, the Central Moment Difference (CMD)~\cite{zellinger2017central} metric is utilized for this purpose. CMD is a sophisticated distance metric that quantifies the dissimilarity between the distributions of two representations by assessing the difference in their sequence moments. The CMD distance diminishes as the distributions become increasingly similar. Consider two bounded random samples, $X$ and $Y$, with probabilities $p$ and $q$, respectively, within the interval $[p, q]$. The central moment difference regularizer is formulated as an empirical estimate of the CMD measure:
\begin{equation}
\begin{aligned}
C M D_{K}(X, Y)=\frac{1}{|b-a|}\|\mathbf{E}(X)-\mathbf{E}(Y)\|_{2}\\
+\sum_{k=2}^{K} \frac{1}{|b-a|^{k}}\left\|C_{k}(X)-C_{k}(Y)\right\|_{2}
\end{aligned}
\end{equation}
In this study, \( \mathbf{E}(X)=\frac{1}{\left | X \right | } \sum_{x\in X}^{} x \) denotes the expectation of the sample vector \( X \), while \( C_k(X)=\mathbf{E}((x-\mathbf{E}(X))^k) \) represents the \( k \)-th central moment of the sample vector \( X \). For the task of Twitter bot detection, this paper computes the CMDloss between invariant representations across two modes.
\begin{equation}
  L_{sim}=\frac{1}{3} \sum_{(m_1,m_2)\in \{(G,T),(G,M),(T,M)\}}^{}CMD_K(h_{m_1}^c,h_{m_2}^c) 
\end{equation}

\subsubsection{Differential Loss $L_{diff}$}
To characterize specific representations between modalities, the loss of difference is minimized to ensure that modalities remain invariant, while specific representations capture distinct aspects of the input and effectively separate the unique information contained in each modality. The distinction between invariant spaces and specific spatial feature representations is achieved by applying soft orthogonality constraints to the two subspaces of each modal representation. In the same batch of trained Twitter users, the matrices \(H_G^i\) and \(H_G^s\) are first normalized to the unit \(L_2\) norm with a zero mean, where the rows represent the hidden vectors \(h_G^i\) and \(h_G^s\) denoted by the graph modal features in the invariant subspace and the specific subspace, respectively. The orthogonality constraint for this modal vector pair is then calculated, and the same transformation is applied to other modalities.
\begin{equation}
  \left \| H_G^iH_G^s \right \| _{F}^{2}  
\end{equation}
In this study, where \( \left \| \cdot \right \| _{F}^{2}  \) represents the square of the Frobenius norm, we introduce orthogonal constraints between vectors in modal-specific subspaces in addition to the existing constraints between vectors in invariant subspaces and those in specific subspaces.
\begin{equation}
\begin{aligned}
L_{diff}=\sum_{m\in \{G,T,M\}}^{} \left \| H_m^iH_m^s \right \| _{F}^{2} + \\
\sum_{(m_1,m_2)\in \{(G,T),(G,M),(T,M)\}}^{}\left \| H_{m_1}^iH_{m_2}^s \right \| _{F}^{2}
\end{aligned}
\end{equation}

\subsubsection{Reconstruction Loss $L_{recon}$}
Despite the introduction of differential losses to mode-specific encoders, there remains a potential risk of learning trivial representations. To address this, our study introduces reconstruction losses to ensure that hidden representations adequately capture the details of their respective modes. Initially, the decoder function \( \hat{x} _m=D(h_m^i+h_m^s;\theta ^{d} ) \) is employed to reconstruct the vector \(x_m\), yielding the vector \( \hat{x} _m \). Subsequently, the mean square error (MSE) loss is computed between \(x_m\) and \( \hat{x} _m \). The loss function, \(\left \| \cdot  \right \| _{2}^{2} \), is defined as the square of the Euclidean norm:
\begin{equation}
  \mathrm{L}_{\text {recon }}=\frac{1}{3}\left(\sum_{m \in\{G, T, M\}} \frac{\left\|\mathbf{x}_{m}-\hat{\mathbf{x}}_{m}\right\|_{2}^{2}}{d_{h}}\right) 
\end{equation}

\subsubsection{Classification Task Loss $L_{task}$}
To optimize the training of BotSAI, this study employs cross-entropy loss to mitigate overfitting. The task loss function is defined as follows:
\begin{equation}
  \mathrm{~L}_{\text {task }}=-\sum_{i \in U}\left[y_{i} \log \left(\hat{y}_{i}\right)+\left(1-y_{i}\right) \log \left(1-\hat{y}_{i}\right)\right]+\lambda \sum_{\theta \in \theta} \omega^{2} 
\end{equation}
Where \( U \) denotes the set of all users in the training dataset, \( \lambda \) is the regularization hyperparameter, \( \theta \) represents all the model's training parameters, \( y_{i} \) denotes the true label of user \( i \), and \( \hat{y}_{i} \) denotes the predicted label of the BotSAI framework for user \( i \).

\section{Experiment}
\label{sec:Experiment}
\subsection{Experimental Setup}

\subsubsection{Dataset.} 
In this study, we examine the impact of various social relationship applications on robot detection outcomes using the well-established and publicly accessible TwiBot-20 and MGTAB datasets. The TwiBot-20 dataset encompasses 229,573 Twitter users from diverse domains such as sports, economics, entertainment, and politics, comprising 33,488,192 tweets. Among these, 11,826 accounts have been identified as either real or automated. The dataset also provides information on the follower and friend relationships between these users.

The MGTAB dataset includes over 1.55 million Twitter users and 130 million tweets, constructing a social network of users with seven types of relationships. For this study, we utilized five of the seven social relationships in the MGTAB dataset: friends, followers, mentions, replies, and references, and labeled 10,199 accounts as either real or bot accounts.

To ensure a fair comparison with previous work, all datasets were randomly divided into training sets, test sets, and validation sets in a 7:2:1 ratio. The training of neural networks inherently involves some randomness, leading to slight variations in model parameters and errors after each iteration, even with fixed hyperparameters and data segmentation. To mitigate the effects of this randomness, the model was trained and tested across five iterations using the same partitioned data. The average performance across these repeated experiments is reported as the final result, thereby smoothing out random fluctuations and providing a more robust assessment of model effectiveness. Using Accuracy and F1-Score as evaluation metrics, experiments were conducted on both datasets.

\subsubsection{Baselines.} 
This paper presents a comprehensive comparative study of BotSAI, considering the following detailed baselines:
\begin{itemize}
\item 
\textbf{Botometer}~\cite{yang2022botometer}: Extracts over 1000 features from user metadata, published content, and user interactions for bot detection. The model is trained using existing datasets.
\item 
\textbf{RoBERTa}~\cite{liu2019roberta}: Utilizes a pre-trained RoBERTa model to encode user text information, which is then input into a multi-layer perceptron (MLP) for bot detection.
\item 
\textbf{SGBot}~\cite{yang2020scalable}: Employs a random forest classifier to extract features from user metadata for scalable and generalized bot detection tasks.
\item 
\textbf{GCN}~\cite{kipf2016semi}: Updates the central node's information by aggregating data from neighboring nodes. It extends the convolution operator to irregular data fields, facilitating connections between graphs and neural networks. This method is widely used for node classification and link prediction.
\item 
\textbf{GAT}~\cite{velivckovic2017graph}: Utilizes a hidden self-attention layer to assign different weights to different neighboring nodes, considering the varying influence of these nodes on the central node. This approach is also widely used for tasks such as link prediction, node classification, and graph clustering.
\item 
\textbf{RGT}~\cite{feng2022heterogeneity}: Employs graph transformers and semantic attention networks to enhance Twitter bot detection within a relational heterogeneous network composed of Twitter users.
\item 
\textbf{BotRGCN}~\cite{feng2021botrgcn}: Constructs a heterogeneous Twitter social network based on user relationships and employs a graph convolutional network for user representation learning and bot detection.
\item 
\textbf{SATAR}~\cite{feng2021satar}: Leverages semantic, attribute, and neighborhood information from Twitter users for user representation learning through a self-supervised approach, followed by fine-tuning on the bot detection dataset.
\item 
\textbf{BIC}~\cite{lei2022bic}: Utilizes a framework of cross-modal information exchange through a text-graphic interaction module and models the semantic consistency of tweets based on attention weights for bot detection tasks.
\item 
\textbf{BotMoE}~\cite{liu2023botmoe}: Integrates various user information, including metadata, text content, and network structure, and combines a community-aware expert hybrid layer (MoE) for robust bot detection and generalization across different Twitter communities.
\end{itemize}

\subsubsection{Experiment Setting.}
\begin{table*}[htbp]
\centering
\begin{tabular}{lcc}
\hline
\textbf{Hyperparameter} & \textbf{TwiBot-20} & \textbf{MGTAB} \\ \hline
Oversampling scale $\omega$ & 0.25 & 1.50 \\ \hline
optimizer & Adam & Adam \\ \hline
learning rate & $10^{-4}$ & $10^{-4}$ \\ \hline
batch size & 128 & 128 \\ \hline
dropout & 0.4 & 0.4 \\ \hline
layer count $L$ & 2 & 2 \\ \hline
hidden size & 256 & 256 \\ \hline
maximum epochs & 400 & 400 \\ \hline
transformer attention head $i$ & 4 & 4 \\ \hline
L2 regularization $\lambda$ & $5 \times 10^{-6}$ & $5 \times 10^{-6}$ \\ \hline
relational edge set $E$ & \{follower, following\} & \{follower, following, reply\} \\ \hline
$\alpha, \beta, \gamma$ & \{0.7, 0.3, 1.0\} & \{0.7, 1.0, 1.0\} \\ \hline
\end{tabular}
\caption{Hyperparameter settings for TwiBot-20 and MGTAB.}
\label{tab:hyperparameters}
\end{table*}

This study employs several software frameworks including Pytorch, Pytorch Geometric, scikit-learn, and Transformers for the implementation of BotSAI. The hyperparameter settings used in the experiments are detailed in Table 1 to facilitate replication. All experiments were executed on a server equipped with an NVIDIA 4060 GPU with 8 GB of VRAM, 16 CPU cores, and 32 GB of RAM. The training durations for BotSAI on the TwiBot-20 and MGTAB datasets were approximately 3 hours and 5 hours, respectively.

\subsection{Main Results}

The initial phase of our evaluation involves determining whether the baseline methods incorporate text data, graphical information, and modal interactions. We then benchmark the TwiBot-20 and MGTAB datasets, demonstrating that BotSAI significantly surpasses all ten feature-based, text-based, and graph-based baselines. The results, detailed in Table 2, reflect the outcomes from five independent runs per method, with averages reported and standard deviations indicated in parentheses. Bold text denotes the highest performance, underlined text indicates the second-highest, and a dash ("-") signifies methods that cannot be extended to MGTAB. Our findings reveal that BotSAI consistently outperforms all baseline methods, including the leading approach, BIC.

1. BotSAI exhibits a significant improvement over all baseline methods across both datasets. Specifically, BotSAI enhances accuracy by 0.81\% on TwiBot-20, improves the F1-score by 1.21\%, achieves a 2.21\% increase on the MGTAB benchmark, and enhances the F1-score by 0.05\% compared to the previous state-of-the-art method, BotMoE, indicating a statistically significant advancement.

2. Experimental results indicate that BotSAI achieves higher detection accuracy on the MGTAB dataset compared to TwiBot-20. This suggests that the strategic utilization of various social relationships in the network (e.g., combinations of friend, follow, and reply) substantially enhances the efficacy of social robot detection.

3. The results show that detection models incorporating multi-modal information, such as BotSAI, BIC, and BotMoE, perform exceptionally well across diverse test environments and datasets. These models significantly outperform traditional single-modal approaches, demonstrating superior adaptability and detection accuracy in the dynamic social media landscape.

4. This study introduces a novel robot detection framework that maps multi-modal information to invariant and specific subspaces for enhanced feature representation. This framework achieves the highest performance in comprehensive benchmark evaluations, underscoring the importance of utilizing multimodal information from Twitter users and validating the effectiveness of the proposed method.

\subsection{Ablation Study}
This paper will investigate the impact of graph data enhancement on detection systems. It will then evaluate the effectiveness of the proposed local relation converter within this context. Subsequently, the paper will analyze the extent to which various social relationships within the Twittersphere influence detection outcomes. Finally, the study will explore the roles of invariant and specific subspaces in the robot detection methodology.
\subsubsection{Graph enhancement}
\begin{figure}[htbp]
  \centering
  \includegraphics[width=\linewidth]{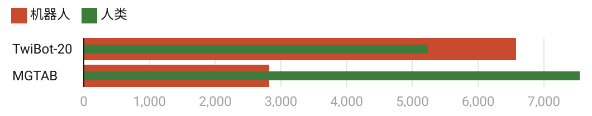}
  \caption{
  The imbalance between the number of genuine users and bot accounts.}
  \label{fig:mix}
\end{figure}
In the TwiBot-20 and MGTAB datasets, users are categorized into robots and humans, revealing an imbalance in their proportions, as illustrated in Figure~\ref{fig:mix}. 
An unbalanced training dataset can pose challenges for models, as they may struggle to learn sufficient features from the minority class, often leading to a bias where the majority class is predominantly predicted. Despite high overall accuracy, these datasets exhibit a significant bias in predictions across most classes, with the classification accuracy for some classes being notably lower~\cite{hu2021graph}. To address this issue, the paper employs an oversampling technique to mitigate the adverse effects of data imbalance during the aggregation of effect information in heterogeneous graphs within the BotSAI framework. Experimental results demonstrate that employing a suitably chosen sampling scale \(\omega \) enhances detection accuracy. These findings suggest that this data augmentation approach effectively reduces the negative impacts of data imbalance and positively contributes to the accuracy of Twitter bot detection.

\subsubsection{Local Relational Transformers}
In the realm of Twitter socialization, bot accounts exhibit a more deliberate behavior compared to genuine users. Advanced bots systematically follow or like accounts within various technology sectors, such as tech bloggers, based on their programmed identities. In contrast, genuine users display a diverse range of interests and exhibit a higher degree of randomness in their interactions. They not only engage with preferred content but also interact with recommendations generated by the system. 

This study introduces a local relation converter that enhances the semantic attention network used in relational graph transformers (RGT). The local relation converter emphasizes the varying significance of different objects within the same social context and incorporates additional weight parameters to more accurately capture user behavior. 
\begin{figure}[htbp]
  \centering
  \includegraphics[width=\linewidth]{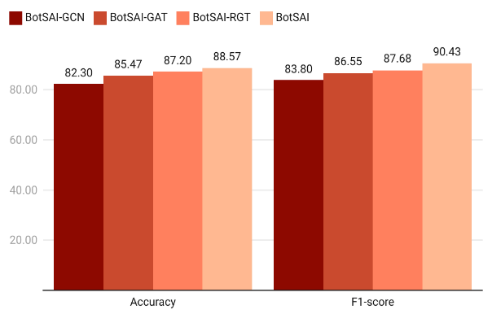}
  \caption{
  Conducting ablation experiments by replacing model components.}
  \label{fig:trans}
\end{figure}

To evaluate the impact of this proposed method on model performance using the Twibot-20 dataset, we conducted ablation experiments by substituting the local relation converter with Graph Convolutional Networks (GCN), Graph Attention Networks (GAT), and RGT, while maintaining consistent parameters. The results, presented in the left panel of Figure~\ref{fig:trans}, indicate that the proposed method outperforms other graph neural network approaches. Specifically, it improves RGT detection accuracy by 1.37\% compared to the most advanced RGT detection technique. This demonstrates that the local relation converter provides superior detailed information aggregation, thereby significantly enhancing detection performance.
\subsubsection{Social Relationship}
There are various methods for socializing within social networks, and advanced bots often simulate genuine interactions by engaging in multiple exchanges with real users. Currently, most detection approaches rely on just two types of social relationships: followers and the followed. However, the cost of employing programmed bots to manipulate these relationships is minimal, enabling significant concealment from genuine users through large-scale manipulation. Consequently, the aggregation of neighboring information from these two relationships can lead to excessive smoothing of bot characteristics, thereby adversely affecting detection accuracy.

To assess the impact of different social relationships on detection performance, this study constructs a social graph by incorporating various relationship combinations within the MGTAB dataset. It evaluates how these relationship settings influence the performance of detection models. The results are detailed in Table 3. Using the BotSAI framework, the heterogeneous social network was modified without altering other parameters for experimentation. Each baseline was executed five times with different seeds, with average performance and standard deviation reported. The most effective results are highlighted in bold, single and compound relationships are underlined, and active and passive relationships are italicized and underlined.

The experimental findings indicate the following:
1. For detecting Twitter bots, a heterogeneous social network that integrates diverse social relationship combinations proves more effective in distinguishing between bots and real users.
2. In ablation experiments focusing on individual social relationships, the accuracy of the Follower group exceeds that of the Following group by 3.12\%. When comparing complex relationships, the accuracy of the Follower group surpasses that of the B+A group by 1.46\%, demonstrating that passive social relations provide more distinctive information between bots and real users compared to active social relations.
3. The detection accuracy of the optimal social relationship combination, A+B+D, is 2.16\% higher than that of the A+B+C+D+E combination, indicating that selecting key social relationships to construct a graph network can mitigate the impact of extraneous noise.

The results underscore that the effectiveness of bot detection heavily relies on the strategic extraction of passive associations. This is attributed to the difficulty that bot manipulators face in concealing passive relationships, as they are more challenging to fabricate compared to active followings. Notably, influential real users tend to attract followers of similar prestige, making it more complex for bots to simulate such accounts. This contrast, as illustrated in the left half of Figure 1, provides robust support for social bot detection efforts.

\subsubsection{Modality invariant Representation}
To investigate the impact of subspace variations on Twitter bot detection, this section examines several variants of the BotSAI model:

BotSAI-BASE serves as the baseline model. In this configuration, three independent encoders are employed for each mode without subspace learning. The encoded features are directly fused, and the resultant vectors are utilized for detection.

BotSAI-SF mirrors BotSAI in the representation learning phase but employs only mode-specific features for fusion and detection.

BotSAI-IF also mirrors BotSAI in the representation learning stage but utilizes only mode-invariant features for fusion and detection.
\begin{figure}[htbp]
  \centering
  \includegraphics[width=\linewidth]{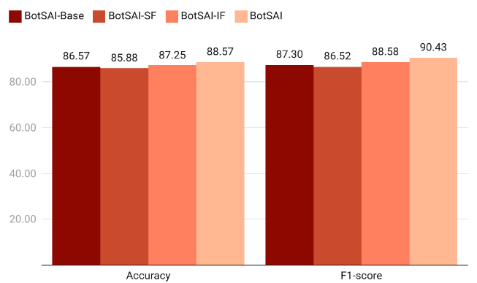}
  \caption{
  Conducting ablation experiments by removing model components.}
  \label{fig:invariant}
\end{figure}

The experimental results, as illustrated in Figure~\ref{fig:invariant}, indicate that the final design proposed in this study outperforms the variants. The BotSAI-SF variant appears to have overfit the differences among the modes, resulting in performance comparable to the BotSAI-BASE model. In contrast, the BotSAI-IF variant demonstrates enhanced detection performance relative to BotSAI-BASE, underscoring the significance of mode-invariant learning. Ultimately, the optimal approach integrates both modal invariance and specific subspace learning to the eigenvector post-representation learning, as demonstrated by the proposed BotSAI model.

\begin{figure}[htbp]
  \centering
  \includegraphics[width=\linewidth]{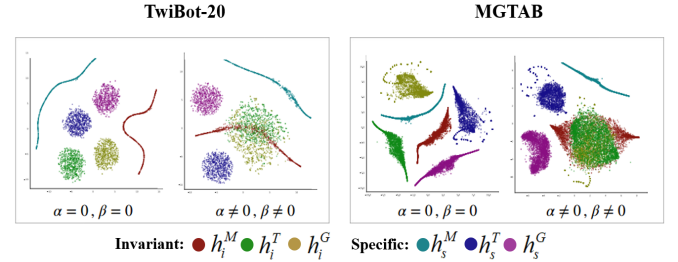}
  \caption{
  Visualization of invariant and specific subspace hidden representations.}
  \label{fig:case}
\end{figure}
To examine the generalization capabilities of invariant and specific subspaces within encoded feature vectors, this study employs T-SNE projection for visualizing the mapping of modal invariant and specific hidden representations in the TwiBot-20 and MGTAB test sets. Figure~\ref{fig:case} illustrates that, in the absence of subspace utilization $(\alpha =0,\beta =0)$, the model fails to capture the commonalities across modes. Conversely, the introduction of losses reveals an overlap between modal invariant representations, demonstrating that BotSAI successfully performs the necessary subspace learning.

\section{Conclusion}
Detecting Twitter bots presents a significant and complex challenge. This paper introduces a novel detection framework, named BotSAI, which is structured into two distinct stages: multi-modal information coding and interaction analysis. The framework projects multi-modal feature vectors onto both invariant and specific subspaces, thereby enhancing the distinct and prominent characteristics across different modes. This approach addresses the challenge posed by sophisticated bots that use camouflage techniques to evade detection. Empirical results demonstrate that BotSAI consistently surpasses current leading methods on two benchmark datasets for Twitter bot detection. Future research will aim to leverage passive social relationships to identify substantial differences between genuine users and bots. This includes the development of datasets that incorporate diverse passive social interactions.

\section{REFERENCES}
\label{sec:refs}

List and number all bibliographical references at the end of the
paper. The references can be numbered in alphabetic order or in
order of appearance in the document. When referring to them in
the text, type the corresponding reference number in square
brackets as shown at the end of this sentence \cite{C2}. An
additional final page (the fifth page, in most cases) is
allowed, but must contain only references to the prior
literature.

\bibliographystyle{IEEEbib}
\bibliography{strings,refs}

\end{document}